\documentclass[prl,twocolumn,notitlepage,superscriptaddress,longbibliography]{revtex4-1}
\usepackage{amsmath}
\usepackage{amssymb}

\usepackage[unicode=true,colorlinks=true,citecolor=blue,urlcolor=blue]{hyperref}

\usepackage{bm}
\usepackage{epsfig}

\usepackage[normalem]{ulem}

\renewcommand {\Im}{\mathop\mathrm{Im}\nolimits}

\renewcommand {\i}{{\rm i}}
\renewcommand {\phi}{{\varphi}}
\newcommand {\rmi}{{\rm i}}
\newcommand {\rmd}{{\rm d}}

\newcommand {\e}{{\rm e}}
\newcommand {\eps}{\varepsilon}

\begin{document}
\title{%
Quantum chaos driven by long-range waveguide-mediated interactions
}

\author{Alexander V. Poshakinskiy}
\affiliation{Ioffe Institute, St. Petersburg 194021, Russia}

\author{Janet Zhong}
\affiliation{Nonlinear Physics Centre, Research School of Physics, Australian National University, Canberra ACT 2601, Australia}

\author{Alexander N. Poddubny}
\email{poddubny@coherent.ioffe.ru}

\affiliation{Ioffe Institute, St. Petersburg 194021, Russia}
\affiliation{Nonlinear Physics Centre, Research School of Physics, Australian National University, Canberra ACT 2601, Australia}

\begin{abstract}
We study theoretically quantum states of a pair of photons interacting with a finite periodic array of two-level atoms in a waveguide. 
Our calculation reveals  two-polariton eigenstates that have a highly irregular wave-function in real space. This  indicates  the  Bethe ansatz breakdown and the onset of quantum chaos, in stark contrast to the conventional integrable problem of two interacting bosons in a box. We identify the long-range waveguide-mediated coupling between the atoms  as the key ingredient of chaos and  nonintegrability. Our results  provide new insights  in the interplay between order, chaos and localization  in many-body quantum systems and can be tested in state-of-the-art setups of waveguide quantum electrodynamics.
\end{abstract}
\date{\today}

\maketitle
{\it Introduction.} 
Arrays of superconducting qubits or cold atoms coupled to a waveguide, have recently become a promising new platform for quantum optics~\cite{Roy2017,KimbleRMP2018,vanLoo2013,Corzo2019,Mirhosseini2019,brehm2020waveguide,Prasad2020}.
They  can be used  for storing ~\cite{Leung2012} and generating quantum light \cite{Zheng2013,Baranger2013,Johnson2019,Prasad2020}, and even  a future ``quantum internet''~\cite{kimble2008quantum}. Moreover, qubit arrays are a new type of quantum simulator for the problems of many-body physics~\cite{Noh2016,Xu2018,Iorsh2020}. One of the most fundamental problems in physics is the  competition between  order and chaos,  or many-body localization and thermalization. It is already a subject of active studies~\cite{Faddeev_2013,Moore2017}, from  celestial mechanics to atomic, nuclear~\cite{Bunakov2016} and   condensed matter ~\cite{Ullmo_2008,DAlessio2016,Assmann2016} physics,  and even  quantum paradoxes in black holes~\cite{Maldacena2016,Morita2019}. Despite the large diversity of these systems, the consideration is typically limited to excitations with parabolic dispersions and short-range coupling.
  Arrays of atoms in a waveguide present a unique platform to probe unexplored boundaries  of quantum chaos and integrability. They offer a special  combination of strong interactions,  long-range waveguide-mediated coupling and intrinsically non-parabolic dispersion of excitations.
\begin{figure}[b]
\includegraphics[width=0.45\textwidth]{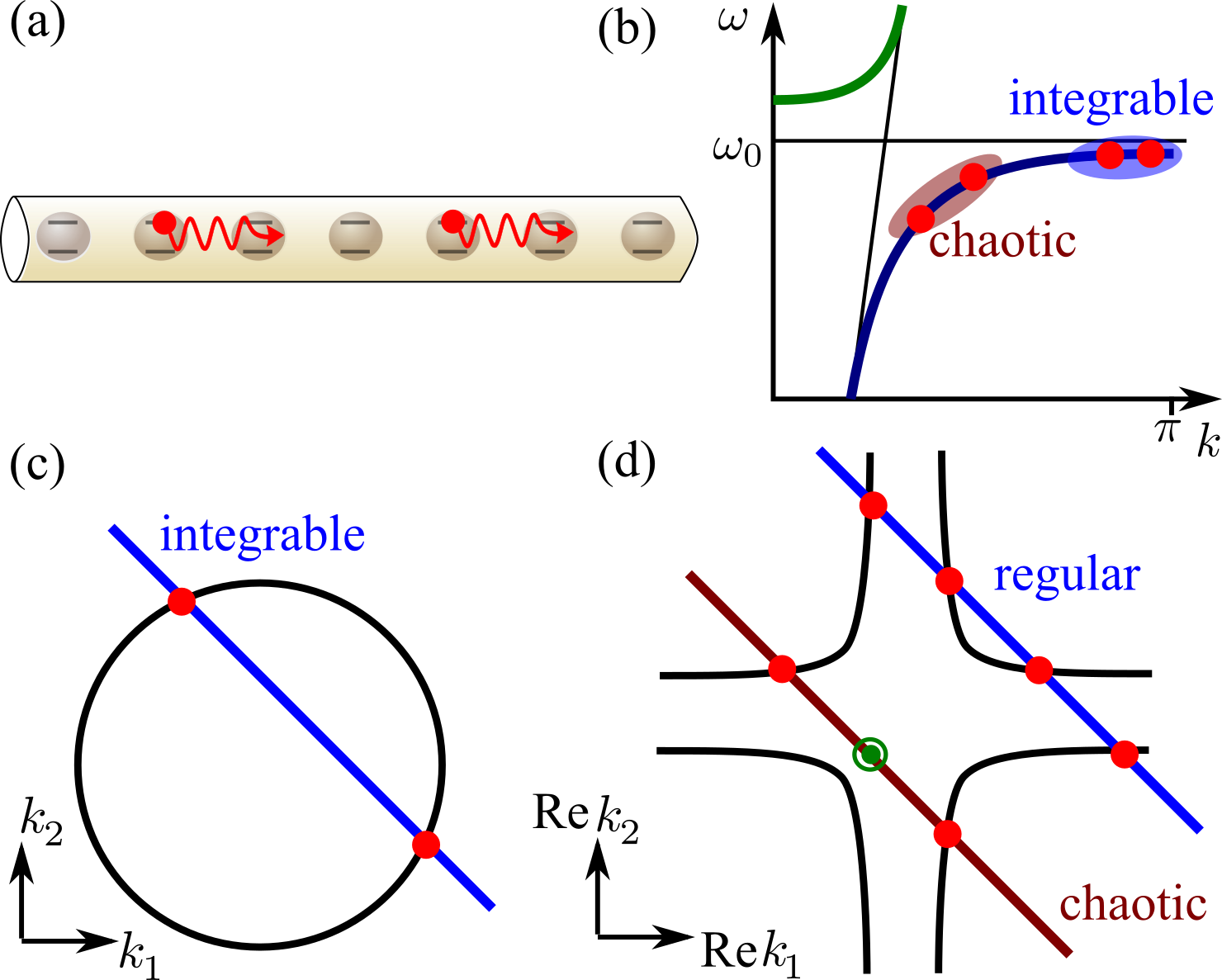}
\caption{(a) Schematics of a two-photon state in array of atoms in a waveguide. (b) Schematics of single-polariton dispersion curve $\omega(k)$. Two polaritons pairs with small and large wave vectors corresponding to chaotic and integrable states are indicated.  (c,d) Wave vectors of two-polariton states with the same total energy and momentum for  the case of (c) parabolic and (d) non-parabolic $\propto -1/k^2$ single-particle dispersion. Black curves shows the isoenergy contour 
$\omega(k_1)+\omega(k_2) =const$. Slanted lines illustrate the total momentum conservation, $k_1+k_2=const$.
Green circle in (c) corresponds to the complex $k_{1,2}$, with  real part outside the isoenergy contour.
}\label{fig:1}
\end{figure}

Here, we consider an interaction of two photons   with a periodic finite array of two-level atoms in a waveguide, illustrated in Fig.~\ref{fig:1}(a). The coupling of photons to atoms leads to the formation of collective polaritonic excitations.  Polaritons repel  each other since a single two-level atom can not host two resonant photons at the same time~\cite{Birnbaum2005}. This is strongly reminiscent of an exactly solvable (integrable) one-dimensional model  of two bosons in a box, that demonstrates fermionization in the limit of strong repulsion~\cite{Lieb1963,McGuire1964,Gaudin1971}. The integrability can be broken when the interaction becomes nonlocal~\cite{Beims2007}, or there is an external potential \cite{Shepelyansky2016}, or if the bosons acquire different masses~\cite{Vessen2001}, which can be mapped to an irrational-angle billiard
~\cite{Casati1999}. Since  considered polaritons are locally interacting  equivalent bosons and there is no external potential the integrability  should persist  at the first glance. Indeed,  fermionized two-polariton states have been  recently revealed by  Zhang and M{\o}lmer ~\cite{Molmer2019}.
However, we later uncovered~\cite{Zhong2020,alex2020quantum} a very different  kind of two-polariton states  that have  a broad Fourier spectrum, and cannot be reduced to a product of several single-particle states. This hints that the problem is non-integrable by the Bethe ansatz.   The mechanism of non-integrability and its possible consequences, such as existence of chaotic two-polariton states remain unclear.
\begin{figure*}[t]
\centering\includegraphics[width=0.85\textwidth]{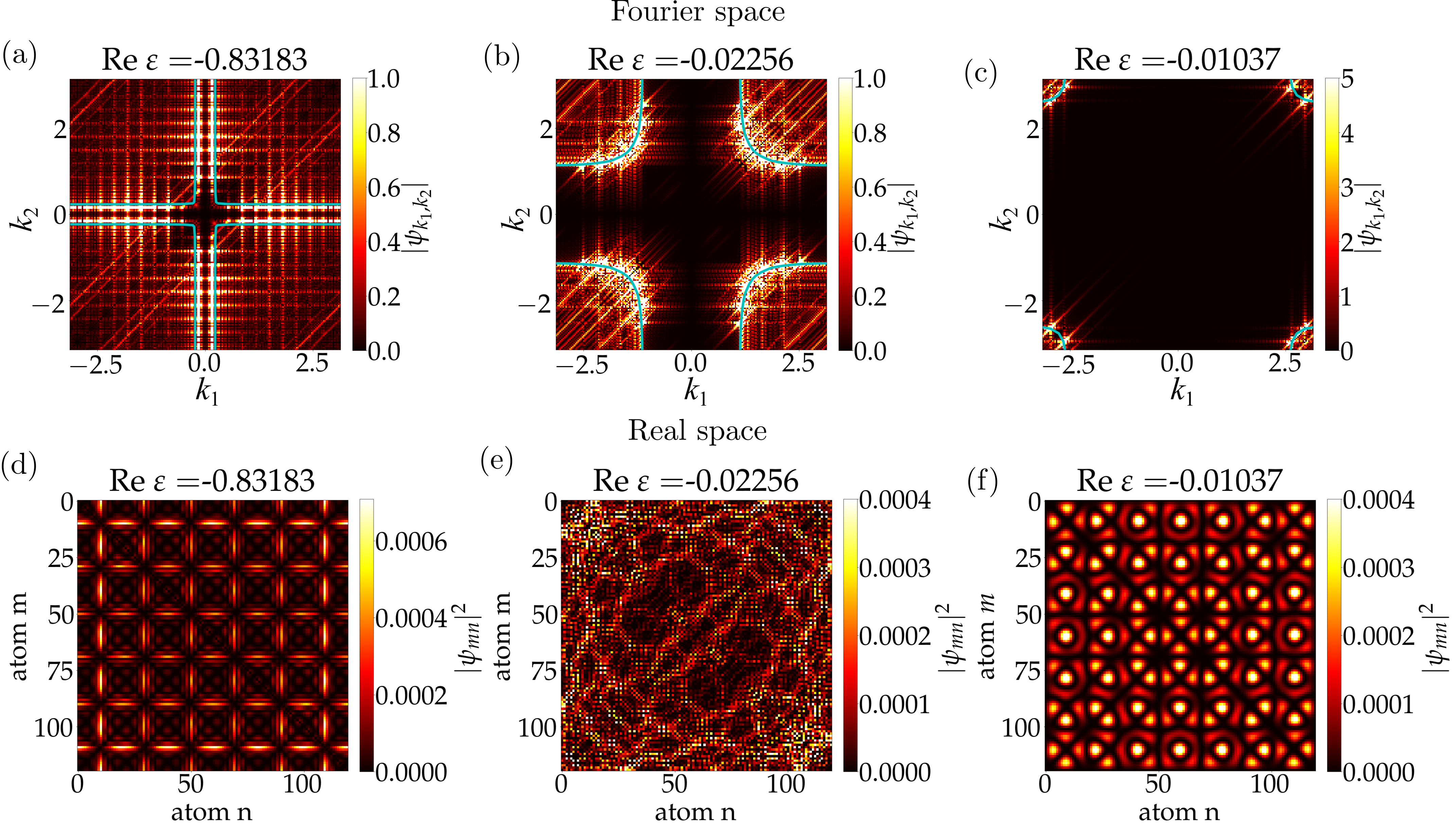}
\caption{(a,b,c) Fourier transforms  and (d,e,f) real-space wavefunctions of several characteristic two-polariton states.
(a,d): regular nonintegrable state, (b,e): irregular chaotic-like state, (c,f): fermionized  state. The calculation has been performed for $N=120$ qubits and $\varphi=0.02$. Cyan curves in (a--c) show the isoenergy contours Eq.~\eqref{eq:iso}. Energy is measured in units of $\Gamma_0$. 
}\label{fig:2}
\end{figure*}

In this Letter, we  examine the transition between the regular two-polariton states \cite{Zhong2020,alex2020quantum} and the fermionized states~\cite{Molmer2019} and identify the emergence of chaotic two-polariton eigenstates at the transition point. 
 In a nutshell, the origin of chaotic states can be understood by analyzing the conservation of energy
 $\omega(k_1)+\omega(k_2) =2\eps$  and center of mass  momentum $k_1+k_2 = K$  for two interacting polaritons, as shown in Fig.~\ref{fig:1}(c,d). In a conventional system with parabolic dispersion  $\omega\propto k^2$ there exist just two pairs of particles with given total energy $2\eps$ and momentum $K$. These two pairs can be found from  the intersection of the isoenergy curve
  $k_1^2+k_2^2= const$ [circle in Fig.~\ref{fig:1}(c)] with the iso-momentum line $k_1+k_2= const$ [blue line in Fig.~\ref{fig:1}(c)]. However,  the dispersion of polaritons is strongly nonparabolic, resulting from avoided crossing of light dispersion $\omega(k)=ck$ with the atomic resonance at $\omega=\omega_0$, see Fig.~\ref{fig:1}(b)~\cite{Ivchenko1991,Albrecht2019,Zhong2020}. Specifically, for the intermediate part of the lower polariton branch away from the Brillouin zone edge one has 
    $\omega(k)\propto -1/k^2$~\cite{Zhong2020} and   the isoenergy curve $\omega(k_1)+\omega(k_2)= const $ acquires a more complicated hyperbolic shape [Fig.~\ref{fig:1}(d)] instead of a circle in Fig.~\ref{fig:1}(c). There exist 4 pairs of polaritons with a given total energy and momentum [blue  line in Fig.~\ref{fig:1}(d)] instead of 2 pairs in Fig.~\ref{fig:1}(c). Moreover, the values of  $k_1$ and $k_2$ can be complex  even when total momentum and energy are real [red line in Fig.~\ref{fig:1}(d)].    We prove below that the combination of polariton-polariton interactions with polariton reflections from  the array edges, when $k_{1,2}\to -k_{1,2}$, makes the number of single-particle states with the same total energy and momentum arbitrarily large. We have found a chaotic nonlinear map that governs the  distribution of  wave vectors $k$ and  thus drives chaotic two-polariton states.
     Such mechanism of  emergence of chaos and nonintegrability is very general and   should apply to various many-body setups with nonparabolic  dispersion of excitations, that is typical for long-range coupling.



{\it Regular and irregular two-photon states}.
We will now present details of the model and numerical results. We consider $N$ periodically spaced qubits in a one-dimensional waveguide, characterized in the Markovian approximation by  the Hamiltonian
 $\mathcal H=\sum_{m,n=1}^NH_{m,n}b_{m}^{\dag}b_{n}+\frac{\chi}{2}\sum_{n=1}^Nb_{n}^{\dag} b_{n}^{\dag}b_{n}^{\vphantom{\dag}}b_n^{\vphantom{\dag}}\:,$
 where
$H_{mn}\equiv-\rmi\Gamma_0\e^{\rmi \varphi |m-n|}\:,\quad m,n=1\ldots N\:.$
Here, $b_{m}$ are the  annihilation operators for the bosonic  excitations of the qubits 
and  $\varphi$ is the phase  acquired by light  between the two neighboring qubits. The details of derivation
can be  found in Refs.~\cite{Caneva2015,Ke2019} and also in Supplementary Materials. The Hamiltonian is non-Hermitian due to the possibility of radiative losses into the waveguide and the coupling strength does not decay with distance.
 We consider subwavelength regime when $\varphi\sim 1/N\ll 1$. The parameter $\Gamma_0$ is the radiative decay rate of an individual qubit and  the anharmonicity  $\chi$ is responsible for polariton-polariton  interactions.
We focus on  the double-excited  states $\sum_{m,n}\psi_{mn} b_n^\dag b_m^\dag |0\rangle$.
In the limit of two-level qubits, when $\chi/\Gamma_0\to\infty$ and $\psi_{nn}\equiv 0$, the  Schr\"odinger equation  for these states reads (see Refs.~\cite{Ke2019,Zhong2020} and Supplementary Materials):
\begin{equation}\label{eq:S2}
H_{nn'}\psi_{n'm}+\psi_{nn'} H_{n'm}-2\delta_{nm} H_{nn'}\psi_{n'n}=2\eps \psi_{nm}\:,
\end{equation}
with    $\psi_{nm}=\psi_{mn}$, and $n,m=1\ldots N$. Here, the first two terms in the left-hand side describe the propagation of  the first and second polaritons, respectively. The third term accounts for their repulsion, enforcing $\psi_{nn}=0$.

Figure~\ref{fig:2} presents three characteristic eigenstates, with the energies increasing from left to right, calculated numerically for an array with $N=120$ qubits. Top row shows two-dimensional Fourier transforms $|\sum_{nm}\psi_{nm}\e^{-\rmi k_x n-\rmi k_x m}|^2$ and the bottom row presents the real-space probability densities $|\psi_{nm}|^2$. 
 The state in Fig.~\ref{fig:2}(a,d) can be understood from the  analytical model  where each one of the two polaritons induces in real space an effective periodic potential for the other one~\cite{alex2020quantum}. It has a regular structure with sharp localized features in  real space,  Fig.~\ref{fig:2}(d)  and  a relatively broad distribution in the Fourier space with many discrete peaks concentrated along the isoenergy contour of non-interacting polariton pair~\cite{Zhong2020},
\begin{equation}
\frac{\Gamma_0\sin\varphi}{\cos k_1-\cos\varphi}+\frac{\Gamma_0\sin\varphi}{\cos k_2-\cos\varphi} = 2\eps \:,\label{eq:iso}
\end{equation}
shown by the cyan curves in Fig.~\ref{fig:2}(a--c).  As such, the  state in Fig.~\ref{fig:2}(a,d)  consists of many single-particle states and clearly cannot be described by a simple Bethe ansatz, although it has a regular real-space wavefunction.
The state in Fig.~\ref{fig:2}(b,e) is very different and we will term it as a chaotic state. While it is hard to give a mathematically precise definition of chaotic states in a finite discrete system, we stress  that the state Fig.~\ref{fig:2}(b) has a highly irregular wavefunction in real space, and, at the same time its Fourier spectrum in  Fig.~\ref{fig:2}(e) is broad and  relatively homogeneous along the isoenergy contour. This is in accordance with the Berry hypothesis for  chaotic states~\cite{Berry1977}. Finally, in  Fig.~\ref{fig:2}(c,d) we show the fermionized two-polariton state~\cite{Molmer2019}. The state is regular in real space, has 8 distinct peaks in the Fourier space, and is well described by the Bethe ansatz
\begin{align}\label{eq:Bethe}
\psi_{nm}=\psi_{mn}\propto &\cos k_1(n-\tfrac{1}{2})\cos k_2(m-\tfrac{1}{2}) \\ \nonumber
-&\cos k_2(n-\tfrac{1}{2})\cos k_1(m-\tfrac{1}{2}) \quad \text{for $n>m$. }
\end{align}
The coexistence  of the fermionized regular eigenstates Fig.~\ref{fig:2}(c,f) with
 regular eigenstates Fig.~\ref{fig:2}(a,d) and chaotic eigenstates Fig.~\ref{fig:2}(b,e) for the same Hamiltonian and the same parameters is rather surprising. Our  central goal is  to explain this result and  to  identify the origin of the apparent chaotic character of the wavefunction 
Fig.~\ref{fig:2}(b,e).

{\it Bethe ansatz and its breakdown}. We first construct  the Bethe ansatz solution for an infinite array and then explain where it fails for a finite array. It is inconvenient to start directly from  the Schr\"odinger equation~\eqref{eq:S2} since the corresponding Hamiltonian matrix is  dense, i.e., includes long-range waveguide-mediated couplings. Instead, we use the fact that the inverse matrix $H^{-1}$ is tri-diagonal,
  and  change the basis as $\psi=H^{-1}\Psi H^{-1}$~\cite{Poddubny2019quasiflat} to obtain an equivalent sparse equation~\cite{Zhong2020}
  \begin{align}\label{eq:main}
(H^{-1}\Psi+ \Psi H^{-1})_{nm}-&2\delta_{nm}(\Psi H^{-1})_{nn} \nonumber\\
&=2\eps (H^{-1}\Psi H^{-1})_{nm}\:.
 \end{align} 
 We  now try to solve it  using a Bethe ansatz 
\begin{equation}\label{eq:ansatz}
\Psi_{mn}= \sum_{K,q} A_{K,q}\,  \e^{\rmi K(m+n)+\rmi q|m-n|/2}
\end{equation}
 where $A_{K,q}$ are the coefficients and  the summation goes over particular values of the center of mass motion wave vector $  K=(k_1+k_2)/2$  and  the relative motion wave vector $q=k_1-k_2$  that are determined below. Each term of the ansatz Eq.~\eqref{eq:ansatz} shall satisfy Eq.~\eqref{eq:main} at all $m,n$ except for the diagonal region  $|m- n|=0,1$ and the array boundaries $m, n = 1, N$. That is fulfilled if $k_{1,2}=K\pm q/2$ lies on the isoenergy contour  Eq.~\eqref{eq:iso}. 

First, we consider an infinite array, where the center of mass wave vector $  K$ is a good quantum number. Substituting $k_{1,2}=K\pm q/2$ in the dispersion equation Eq.~\eqref{eq:iso} we find 4 inequivalent values of the relative motion wave vector $q(K)$ for any value of $K$. 
 The values of $q$   can be both real and complex, explicit expressions are given in the Supplementary Materials. Real-valued solutions can be found from  the intersection of the line $k_1+k_2=K$, describing all states with given total momentum, with the isoenergy contour Eq.~\eqref{eq:iso}, see Fig.~\ref{fig:1}(d).
  These four solutions  can be combined in Eq.~\eqref{eq:ansatz} to satisfy Eq.~\eqref{eq:main} as shown in the Supplementary Materials which  finishes the construction of the Bethe ansatz in the infinite system.   However,  this procedure breaks down for a finite array.
\begin{figure}[b]
\includegraphics[width=0.48\textwidth]{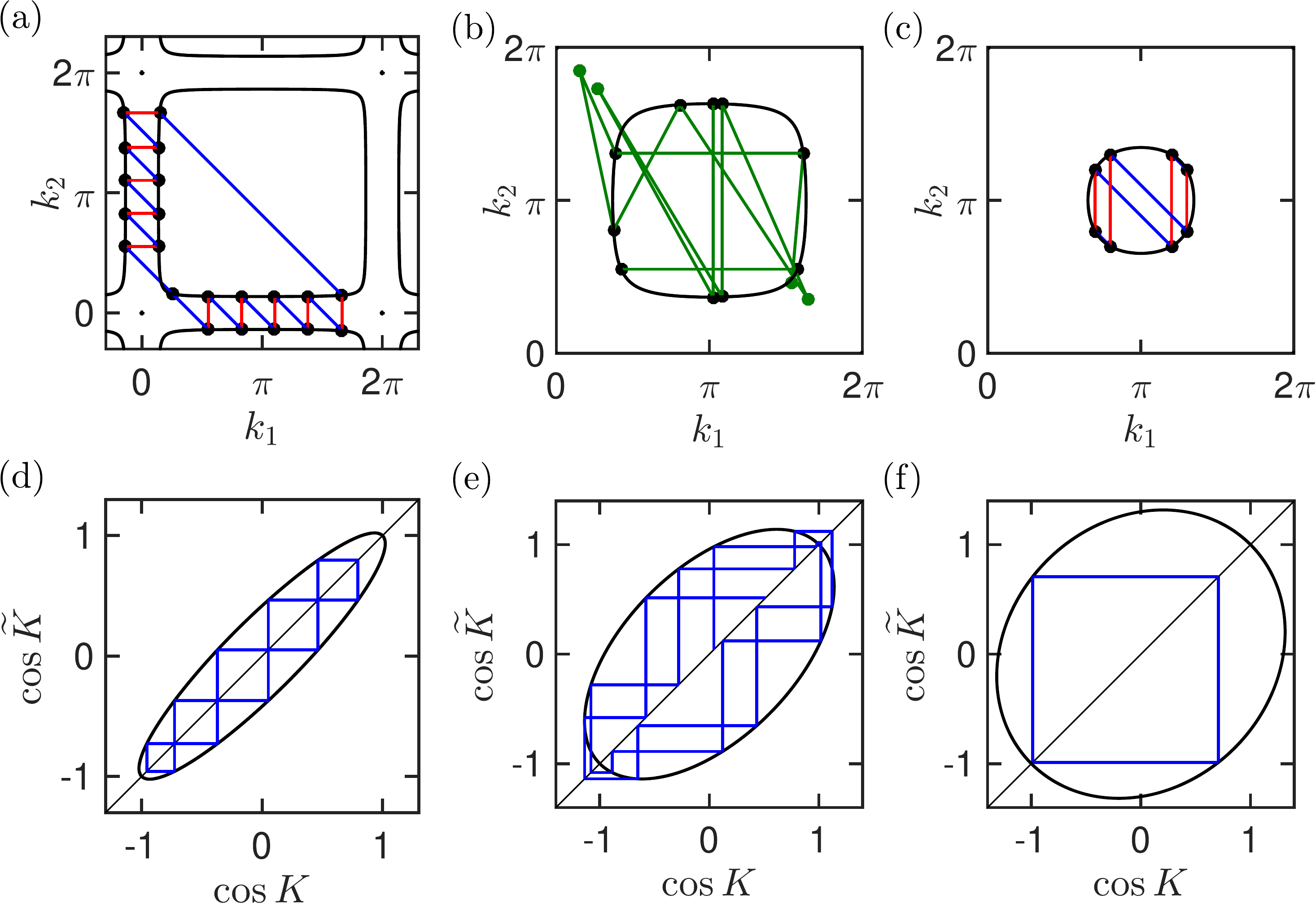}
\caption{
Examples of subsequent application of the map Eq.~\eqref{eq:map}, for different initial wave vectors:
(a) regular 21-cycle, starting with {$k_1=0.8$, $k_2=0.5$},
(b) ergodic infinite cycle starting with 
$k_1=1.33$, $k_2=1.73$, (c) 8-cycle in  fermionic regime starting with {$k_1=2.2$, $k_2=3.8$}.
Green points in (b) show real parts of  complex wave vectors.
 For (b,c) the vectors $k_{1,2}$ are reduced to the Brillouin zone $0\le k_{1,2}\le 2\pi$ before the vector $K=(k_1+k_2)/2$ is calculated.
Black lines show the isofrequency contours Eq.~\eqref{eq:iso}.
Bottom panels (d-f) show the same cycles as in (a-c) but plotted for the equivalent map
Eq.~\eqref{eq:map1}, tracing the evolution of the center-of-mass wave vector $K$. 
}\label{fig:3}
\end{figure}

In a finite array, photons can reflect from the boundaries. To  accommodate  the boundaries, one should include in Bethe ansatz the reflected waves with the wave vectors $\widetilde k_{1,2}= -k_{1,2}$. After the reflection of one of the two photons, the new center of mass wave vector is  $\widetilde K=(\widetilde k_1+\widetilde k_2)/2=\pm (k_1-k_2)/2=\pm q/2$. Thus, we obtain a nonlinear map
\begin{equation}\label{eq:map}
K\to \widetilde K=\pm \frac{1}{2}q(K)\:,
\end{equation} 
which generates new pairs of wave vectors $K$ and $q(K)$ that 
 must be included into the Bethe ansatz Eq.~\eqref{eq:ansatz}. All the generated plane waves 
 should be  combined to satisfy the Schr\"odinger equation at the boundaries~\cite{Batchelor2007,Kollar2012}.  
 The impossibility to do so would indicate that the system is non-integrable.
However, the considered two-polariton problem offers one more scenario of the Bethe ansatz breakdown. Namely, the map Eq.~\eqref{eq:map} can generate an arbitrarily large number of wave vectors, rendering the whole Bethe ansatz construction  impractical. 

In three columns   Fig.~\ref{fig:3},  we will now explore the map for different  ranges of wave vectors $k_{1,2}$ that feature regular, chaotic and fermionic two-polariton states.  
We start with Fig.~\ref{fig:3}(a) that corresponds to the situation of 
Fig.~\ref{fig:2}(a), where $k_1\ll \pi; k_2\gg k_1$ and the isoenergy contour is almost flat. The subsequent reflections (red lines) and the map $q(K)$ evaluation 
(intersection of the  isoenergy contour with the blue lines $k_1+k_2=\rm const$) yield two  ``chainsaws''  of almost equidistant points.  Figure~\ref{fig:3}(a) shows a specific cycle  with just 21 points, but the length of cycle can be arbitrarily large. The set of wave vectors obtained  in Fig.~\ref{fig:3}(a) explains the Fourier transform of wavefunction in Fig.~\ref{fig:2}(a).
It is instructive  to rewrite the  map Eq.~\eqref{eq:map} as a quadratic form depending on
$\cos K$ and $\cos \widetilde K$. For $\varphi\ll 1$ the map can be presented as 
\begin{equation}
(\cos K-\cos \widetilde K)^2-\frac{\varphi \Gamma_0}{\eps}(\cos K\cos\widetilde K-1)=0\:.\label{eq:map1}
\end{equation}
Figure~\ref{fig:3}(d) shows the same iterations as Fig.~\ref{fig:3}(a) for the $K\to \widetilde K$ map Eq.~\eqref{eq:map1}.

Another scenario is realized   when $k_1$ and $k_2$ are  both close to the Brillouin zone edge $\pi$. The polariton dispersion  is then almost parabolic~\cite{Molmer2019} and the isoenergy contours \eqref{eq:iso} reduce to slightly deformed circles centered at $k_{1,2}= \pm \pi$, see Fig.~\ref{fig:3}(c,f). As such,  the map Eq.~\eqref{eq:map} generates just 8 inequivalent points, similar to the traditional Bethe ansatz~\cite{Longhi2013}.  This  explains the  fermionic states~\cite{Molmer2019}, shown in Fig.~\ref{fig:2}(c). However, this consideration fails for intermediate values of wave vectors since it takes into account only two real values of $q$ for each center of mass wave vector and ignores two other (complex) values. 
%
\begin{figure}[t]
\includegraphics[width=0.49\textwidth]{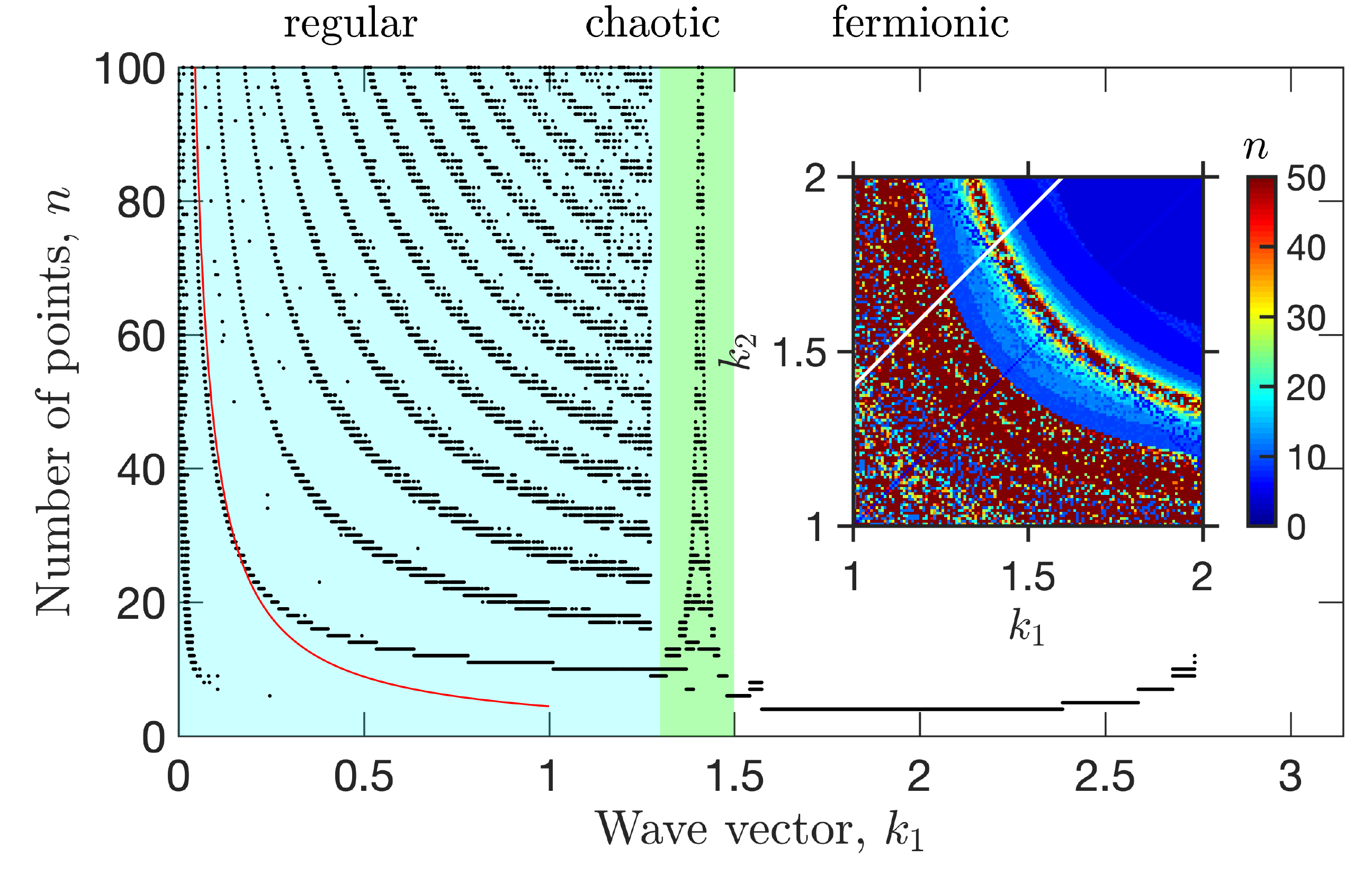}
\caption{Number of points $n$ generated by the map Eq.~\eqref{eq:map} depending on the starting wave vector $k_1$
for $k_2=k_1+0.4$. We used 100 iterations for each of the $1.6\times 10^5$ starting values of $k_1=0.4\ldots \pi-0.4$ (white line in the inset). 
Only the points below the threshold $|\Im q|<1$ have been included in the map. Inset shows the dependence of $n$ on both starting wave vectors $k_1$ and $k_2$ varying  near the center of the Brillouin zone. The grid step is $8\times 10^{-3}$ and 70 iterations were made for the inset.
}\label{fig:4}
\end{figure}
%
When the evanescent waves with complex $q,K$ are taken into account, the maps  Eq.~\eqref{eq:map},\eqref{eq:map1} can generate  infinite ergodic trajectories. In order to build ergodic trajectory we use  the fact that  the map Eq.~\eqref{eq:map1} provides two values of $\cos \widetilde K$ for each value of $\cos K$. By choosing  between these two values we can build  an infinite trajectory that turns around  the points $\cos K=\cos \widetilde K=\pm 1 $ and never repeats itself, as shown in    Fig.~\ref{fig:3}(e).   By construction, this trajectory  includes evanescent waves, where $|\cos K|>1$ and the polariton wave vectors $k_{1,2}=K\pm q/2$ are complex. This is  also seen in Fig.~\ref{fig:3}(b), where  green points represent complex $k_{1,2}$ that  do not lie on the real isoenergy contour.  Such trajectories lead to a dense irregular distribution of wave vectors in the Fourier space and explain formation of chaotic states  Fig.~\ref{fig:2}(b,e).  

In order to examine the transition from  regular to chaotic states in more detail we plot in Fig.~\ref{fig:4} the number of points generated by the  map Eq.~\eqref{eq:map} depending on the initial polariton wave vector $k_1$ for $k_2=k_1+0.4$. 
Three distinct ranges of wave vectors can be identified. In the range $0\le k_1\lesssim 1.3$ the map generates  cycles of type Fig.~\ref{fig:3}(a,d). The points in  Fig.~\ref{fig:4} group into ``lines'' that correspond to cycles with different number of loops made around the ellipse in Fig.~\ref{fig:3}(d). For example, the red curve $n=\sqrt{2}\pi/k_1$ shows the approximate number of points for  a one-loop cycle. 
In our calculation we neglected strongly evanescent waves with  $|\Im q|> \Im q^*=1$ assuming that their contribution to the wave function is exponentially weak. Such cutoff leads to a steep decrease of the number of  generated points  for  $k_1\gtrsim 1.5$ (the results are not qualitatively sensitive to the cutoff value).  Only a small number of wave vectors are generated, which corresponds to the  fermionized states of the type Fig.~\ref{fig:3}(c,f). 
 Finally, there is a narrow peak in the transition region, centered at around $k_1\approx 1.4$, corresponding to the  chaotic states  of the type Fig.~\ref{fig:3}(b,e).  
Inset of Fig.~\ref{fig:4} shows the same number of generated points depending on the values of both initial wave vectors $k_1$ and $k_2$. The calculation also reveals two distinct regions of fermionic and regular states, with a narrow chaotic region in between. 

To summarize, we have obtained a nonlinear map describing two-polariton interactions in $k$-space. The number of non-evanescent waves generated by this map is a good predictor whether a given quantum state is regular non-integrable (small values of $k_{1,2}$), chaotic (intermediate values of $k_{1,2}$) or integrable fermionized ($k_{1,2}$ close to the edge of the Brillouin zone). Our findings apply  to various two-particle systems  and will be hopefully useful also for the many-body setups. Experimental verification  could be  done with already available  arrays of tens of superconducting qubits ~\cite{kim2020quantum,brehm2020waveguide} with the possibility to excite and probe every qubit separately \cite{Ye2019}.


\let\oldaddcontentsline\addcontentsline
\renewcommand{\addcontentsline}[3]{}


%

\newpage

\let\addcontentsline\oldaddcontentsline

\newpage

\setcounter{figure}{0}
\setcounter{section}{0}
\setcounter{equation}{0}
\renewcommand{\thefigure}{S\arabic{figure}}
\renewcommand{\thesection}{S\Roman{section}}
\renewcommand{\thesection}{S\arabic{section}}
\renewcommand{\theequation}{S\arabic{equation}}
\appendix
\begin{center}
\textbf{\Large Online Supplementary Materials}
\end{center}
\tableofcontents
\section{Derivation of the two-polariton  Hamiltonian}
In this section we provide some details on the  derivation of the two-polariton Schr\"odinger equation Eq.~(1) in the main text. 
The derivation follows Supplementary Materials of  Refs.~\cite{Ke2019,Zhong2020}, alternative but equivalent derivations can be found in Refs.~\cite{Caneva2015,Zhang2019arXiv}.

We start with the  Hamiltonian for interaction between array of atoms and photons
 \begin{multline}\label{eq:H0}
 \mathcal  H=\sum\limits_{k}\omega_{k}a_{k}^{\dag}a_{k}^{\vphantom{\dag}}
 +\sum\limits_{j}\omega_{0}b_{j}^{\dag}b_{j}^{\vphantom{\dag}}+\frac{\chi}{2}
 \sum\limits_{j}b_{j}^{\dag}b_{j}^{\dag}b_{j}^{\vphantom{\dag}}b_{j}^{\vphantom{\dag}}
 \\+\frac{ g}{\sqrt{L}}\sum\limits_{j,k}
 (b_{j}^{\dag}a_{k}\e^{\rmi k z_{j}}+b_{j}a_{k}^{\dag}\e^{-\rmi k z_{j}})\:.
 \end{multline}
Here, $a_{k}$ are the annihilation operators for the waveguide photons  with the wave vectors $k$, frequencies $\omega_{k}=c|k|$ and the velocity $c$, $g$ is the interaction constant, $L$ is the normalization length, and $b_{j}$ are the (bosonic) annihilation operators for the qubit excitations with the frequency $\omega_{0}$, located at the point $z_{j}$. 
In Eq.~\eqref{eq:H0}, we consider the general case of anharmonic many-level qubits.  The two-level case can be obtained in the limit of large anharmonicity ($\chi \to \infty$) where the multiple occupation is suppressed~\cite{Baranger2013,Poshakinskiy2016,abrikosov1965electron}.
The photonic degrees of freedom can be integrated out in Eq.~\eqref{eq:H0} yielding the effective Hamiltonian \cite{Caneva2015,Ivchenko2005}
\begin{equation}  H_{mn}=\omega_0\delta_{nm}-\rmi \Gamma_0 \e^{\i \omega |z_m-z_n|/c} \end{equation}
 describing the motion of qubit excitations
\begin{eqnarray}\label{eq:Hmn1}
  H_{mn}&=&\omega_0\delta_{m,n}+g^2 \sum_{l,l'}\int\frac{\rmd k }{2\pi}\e^{\i (z_l-z_l')}\frac{\langle 0|b_m a_k b_l^\dag b_l' a_k^{\dag}  b_n^\dag|0\rangle }{\omega-\omega_k+{0^+}\i} \nonumber \\
  &=&\omega_0\delta_{m,n}+g^2 \sum_{l,l'}\int\frac{\rmd k }{2\pi}\frac{\e^{\i (z_m-z_n)}}{\omega-c|k|+{0^+}\i} \nonumber \\
  &=& \omega_0\delta_{m,n}-\i\frac{g^2}{c}\e^{\i \omega |z_m-z_n|/c}\:.
\end{eqnarray}
Here we have introduced the radiative decay rate  $\Gamma_0={g^2}/{c}$ and implied the rotating wave approximation.
From now on we will count the energy from $\omega_0$ and hence omit  the $ \omega_0\delta_{m,n}$ term.
Then, the total effective Hamiltonian is given as
\begin{align}
\mathcal H=\sum\limits_{m,n=1}^NH_{m,n}(\omega_{0})b_{m}^{\dag}b_{n}+\frac{\chi}{2}
\sum\limits_{m=1}^Nb_{m}^{\dag}b_{m}^{\dag}b_{m}^{\vphantom{\dag}}b_{m}^{\vphantom{\dag}}\:.
\end{align}
Here we use the Markovian approximation, by replacing the phase  $\omega |z_m-z_n|/c$ in Eq.~\eqref{eq:Hmn1} by $\omega_0 |z_m-z_n|/c$.
When being limited to the subspace with only two excitations,  we can construct the effective two-photon Hamiltonian
\begin{equation}
H^{(2)}_{i_{1}i_{2};j_{1}j_{2}}=\delta_{i_{2},j_{2}}H^{(1)}_{i_{1}j_{1}}+\delta_{i_{1},j_{1}}H^{(1)}_{i_{2}j_{2}}+U_{i_{1}i_{2};j_{1}j_{2}}
\end{equation}
where $i_{1},i_{2},j_{1},j_{2}=1\ldots N$ and 
\begin{equation}
\mathcal U_{i_{1}i_{2};j_{1}j_{2}}=\delta_{i_{1}i_{2}}\delta_{j_{1}j_{2}}\delta_{i_{1}j_{1}}\chi\:.
\end{equation}
The linear eigenvalue problem to obtain the two-particle excitations then reads
\begin{gather}\label{eq:S1}
H_{mn'}\psi_{n'n}+\psi_{mn'}H_{n'n}+\chi \delta_{mn}\psi_{nn}=2\eps \psi_{mn}
\end{gather}
We now proceed  to the limit of two-level atoms, when $\chi\to\infty$.
Importantly, even though $\psi_{nn}\to 0$ for $\chi\to\infty$, we still have
 $\chi \psi_{nn}\to\rm const$.
 The value of  $\chi \psi_{nn}$ for large $\chi$ can be calculated perturbatively
\begin{multline}
\chi \psi_{nn}=-\sum\limits_{n'\ne n}(H_{nn'}\psi_{n'n}+\psi_{nn'}H_{n'n})=\\-
2\sum\limits_{n'\ne n}H_{nn'}\psi_{n'n}=-2\sum\limits_{n'=1}^{N}H_{nn'}\psi_{n'n}\:.
\end{multline}
Hence, we can rewrite the Schr\"odinger equation in the limit $\chi\to \infty$ as
\begin{gather}
H_{mn'}\psi_{n'n}+\psi_{mn'}H_{n'n}-2\delta_{mn}H_{nn'}\psi_{n'n}=2\eps \psi_{mn}
\end{gather}
in agreement with Eq.~(1) in the main text.

\section{Dispersion equation}
Here we provide the details of the derivation of the  dispersion equation for the relative motion of two interacting polaritons in the center of mass reference frame. We start from the Schr\"odinger equation  Eq. (1) in the main text, that reads~\cite{Ke2019}
  \begin{align}\label{eq:main1}
(H^{-1}\Psi+ \Psi H^{-1})_{nm}-&2\delta_{nm}(\Psi H^{-1})_{nn} \nonumber\\
&=2\eps (H^{-1}\Psi H^{-1})_{nm}\:.
 \end{align} 
The inverse of the  matrix $H_{mn}=-\rmi \Gamma_0\e^{\rmi \varphi|m-n|}$ is a tri-diagonal matrix \cite{Poddubny2019quasiflat} that explicitly reads 
\begin{equation}\label{eq:iH}
[H^{-1}]_{rs}=\frac{1}{\Gamma_0}\begin{pmatrix}
 -\frac{1}{2}\cot \phi+\frac{\rmi}{2}&\frac1{2\sin\phi}&0&\ldots\\
 \frac1{2\sin\phi}&-\cot \phi&\frac1{2\sin\phi}&\ldots\\
  &&\ddots&\\
  \ldots& \frac1{2\sin\phi}&-\cot \phi& \frac1{2\sin\phi}\\
 \ldots  &0& \frac1{2\sin\phi}& -\frac{1}{2}\cot \phi+\frac{\rmi}{2}
 \end{pmatrix}\:.
 \end{equation} 
  Due to the translation symmetry of the infinite array the two-polariton wavefunction can be sought  in the form 
 \begin{equation}\label{eq:psi-com}
 \Psi_{mn}=\e^{\rmi K(m+n)}\psi_{r},\quad r=|m-n|
  \end{equation} 
  where $K$ is the center of mass wave vector and $\psi_r=\psi_{-r}$ due to bosonic symmetry.
Substituting Eq.~\eqref{eq:psi-com} into the Schr\"odinger equation \eqref{eq:main1} we obtain the 
equations the wavefunction $\psi_r$ that describes relative motion of the two interacting polaritons. The advantage of the equation Eq.~\eqref{eq:main1} based on the inverse Hamiltonian matrix over the center-of-mass motion equation
in Supplementary Materials of \cite{Molmer2019} is that it includes only nearest-neighbor couplings.
Hence, for     $r=|n-m|\ge 2$ we obtain a conceptually simple tight-binding equation
  \begin{multline} \label{eq:r2}
\frac{\cos K}{\sin\phi}(\psi_{r-1}+\psi_{r+1})-2\cot \phi \psi_r\\=  \frac{\eps}{2\Gamma_0\sin^2\phi}[(4\cos^2\phi+2\cos 2K)\psi_r\\+\psi_{r-2}+\psi_{r+2}-4\cos\varphi\cos K(\psi_{r-1}+\psi_{r+1})]\:.
      \end{multline} 
      The values of the relative distance  $r=0$ and $r=1$ are special because one should take into account include non-zero contributions from the polariton-polariton interaction term
$ 2\delta_{nm}(\Psi H^{-1})_{nn} $ in Eq.~\eqref{eq:main1}. Specifically, for   $r=0$ we find 
  \begin{equation}
   \left( 2\cos^{2}\phi+\cos 2K\right) \psi_{{0}}-4\,\cos  \phi
\cos K \psi_{{1}}+\psi_{2}=0\:.\label{eq:r0}
    \end{equation} 
    and    for $r=1$ the Schr\"odinger equation reads
      \begin{multline} 
\frac{\cos K}{\sin\phi}(\psi_0+\psi_2)-2\cot \phi \psi_1\\=  \frac{\eps}{2\Gamma_0\sin^2\phi}[(4\cos^2\phi+2\cos 2K+1)\psi_1\\+\psi_{3}-4\cos\varphi\cos K (\psi_{0}+\psi_{2})] \:.\label{eq:r1}
      \end{multline} 
          For $r\ge 2$ we can use the following ansatz in Eq.~\eqref{eq:r2}
      \begin{equation}
      \psi_r=\e^{\rmi qr/2}
       \end{equation}
       which leads to
       \begin{equation}
      2\eps=\frac{\Gamma_0\cos\phi}{\cos  k_1-\cos\phi}+\frac{2\Gamma_0\cos\phi}{\cos k_2-\cos\phi}\label{eq:eps1}
       \end{equation}      
where $k_{1,2}=K\pm q/2$, which is Eq.~(2) in the main text. This is the presentation of the  total pair energy $2\eps$
is given by a sum of energies of non-interacting polaritons with the wave vectors $k_1$ and $k_2$.
       \renewcommand{\epsilon}{\varepsilon}
       It is more convenient to rewrite the dispersion equation Eq.~\eqref{eq:eps1} as 
       \begin{multline}\label{eq:z}
       -\epsilon\,({z}^{4}+1)+ 2\left( 2\cos  \phi \epsilon+
\sin \phi \right) \cos K({z}^{3}+z)\\-
 2\left(2\cos^{2}\phi\epsilon+
\epsilon \cos 2K+\sin2\phi  \right) {z}^{2}=0\:,
              \end{multline}   
              where $z=\e^{\rmi q/2}$.  The representation Eq.~\eqref{eq:z} explicitly shows  that there are four inequivalent solutions $z_1,1/z_1,z_2,1/z_2$ for each value of total energy of two polaritons $2\eps$ and center of mass wave vector $K$. Dividing Eq.~\eqref{eq:z} by $z^2$ and using the relation $z+1/z=2\cos \frac{q}{2}$ we find for $\varphi\ll 1$
 \begin{equation}
\left(\cos K-\cos \frac{q}{2}\right)^2-\frac{\varphi \Gamma_0}{\eps}(\cos K\cos\frac{q}{2}-1)=0\:.\label{eq:map1b}
\end{equation}
which is equivalent to the map Eq.~(7) in the main text.

Explicit expressions for the wave vectors $q$ can be most easily obtained  for $\varphi\ll 1$ when 
$\omega(k)\approx -2\varphi/k^2$ and 
Eq.~\eqref{eq:eps1} simplifies to
        \begin{equation}
               \eps=-\frac{\Gamma_0\phi}{k_1^2}-\frac{\Gamma_0\phi}{k_2^2}\:.
\end{equation}      
Solution of this equation for $q$ vs $K$ yields 
 \begin{align}
 q_{1,\pm}^2&=\pm \sqrt{ K^2-\frac{4}{w}-\frac{4\sqrt{1- K^2w}}{w}},\\
  q_{2,\pm}^2&=\pm \sqrt{ K^2-\frac{4}{w}+\frac{4\sqrt{1- K^2w}}{w}},
 \end{align}
where  $\eps=w\phi \Gamma_0$.
 
\section{Bethe ansatz}
In this section we provide more details  on the construction of Bethe ansatz in the infinite array. The idea behind this construction is to present the   two-polariton wavefunction as a superposition of single-polariton states and then to satisfy Eqs.~\eqref{eq:r0},\eqref{eq:r1} describing polariton-polariton interactions.
We start with  a general Bethe ansatz expansion
 \begin{equation}\label{eq:Bethe1}
  \psi_m(K,\eps)=\sum\limits_{\nu=1}^{4} \e^{\rmi q_{\nu}m/2}A_\nu
  \end{equation}
  that presents the two-polariton wavefunction as a superposition of solutions with given center-of-mass wave vector $K$ and four possible values of relative motion wave vectors $q$. Substituting Eq.~\eqref{eq:Bethe1} in Eq.~\eqref{eq:r0} we find
\begin{multline}\label{eq:1st}
\sum\limits_{\nu=1}^{4}  A_\nu \bigl[   \left( 2\cos^{2}\phi+\cos 2K\right) \\-4\,\cos  \phi
\cos K\e^{\rmi q_\nu/2}+\e^{\rmi q_\nu}\bigr]=0\:.
  \end{multline}
The same procedure for  Eq.~\eqref{eq:r1} 
  \begin{equation}
  \sum\limits_{\nu=1}^{4}  A_\nu \sin \tfrac{q_\nu}{2}=0\:.\label{eq:2nd}
    \end{equation}
    Solution of Eqs.~\eqref{eq:1st},\eqref{eq:2nd} allows us to express  two of the $A$ coefficients vs other two ones.
Taking into account that  the four wave vectors $q_\nu$ come into two pairs $q_A,-q_A$ and $q_B,-q_B$  , we can rewrite  
Eq.~\eqref{eq:Bethe1} as 
    \begin{multline}
      \psi_m(K,\eps)=(A\e^{\rmi q_Am/2}+\widetilde A\e^{-\rmi q_Am/2})\\+(B\e^{\rmi q_Bm/2}+\widetilde B\e^{-\rmi q_Bm/2})
    \end{multline}
and find 
    \begin{align}
    \widetilde A=&\frac{\sin\frac{q_A}{2}f^*(q_B)+\sin\frac{q_B}{2}f(q_A)}{\sin\frac{q_A}{2}f^*(q_B)\sin\frac{q_B}{2}f^*(q_A)}A\\+&
    \sin\frac{q_B}{2}\frac{f^*(q_B)+f(q_B)}{\sin\frac{q_A}{2}f^*(q_B)-\sin\frac{q_B}{2}f^*(q_A)}B\:,\nonumber\\
        \widetilde B=&-\frac{\sin\frac{q_A}{2}f(q_B)+\sin\frac{q_B}{2}f(q_A)^*}{\sin\frac{q_A}{2}f^*(q_B)-\sin\frac{q_B}{2}f^*(q_A)}B\\-&
    \sin\frac{q_A}{2}\frac{f^*(q_A)+f(q_A)}{\sin\frac{q_A}{2}f^*(q_B)-\sin\frac{q_B}{2}f^*(q_A)}A\:,\nonumber
        \end{align}
        where
    \begin{equation}
    f(q)= \left( 2\cos^{2}\phi+\cos 2K\right) -4\,\cos  \phi
\cos  K\e^{\rmi q/2}+\e^{\rmi q}\:.
    \end{equation}

\end{document}